\documentclass[aps,pra,twocolumn, superscriptaddress]{revtex4-2}
\usepackage{amsmath,amsfonts,amsthm,bbm}
\usepackage[dvipsnames]{xcolor}
\usepackage{graphicx}
\usepackage{multirow,mhchem,chemformula}
\usepackage[colorlinks=true,breaklinks=true,linkcolor=red,citecolor=magenta,urlcolor=magenta]{hyperref}

\usepackage{tikz}
\usetikzlibrary{quantikz}
\usepackage [english]{babel}
\usepackage [autostyle, english = american]{csquotes}
\MakeOuterQuote{"}
\usepackage[colorinlistoftodos]{todonotes}
\usepackage{subfigure}
\usepackage{float}
\usetikzlibrary{quantikz}

\begin{document}

\title{Simulation of a Diels-Alder Reaction on a Quantum Computer}

\author{Ieva Liepuoniute*}
\affiliation{IBM Quantum, IBM Research - Almaden, 650 Harry Road, San Jose, CA 95120, USA}
\thanks{Corresponding authors: ieva@ibm.com, gojones@us.ibm.com}
\author{Mario Motta}
\affiliation{IBM Quantum, IBM Research - Almaden, 650 Harry Road, San Jose, CA 95120, USA}
\address{IBM Quantum, T. J. Watson Research Center, Yorktown Heights, NY 10598, USA}
\author{Thaddeus Pellegrini}
\address{IBM Quantum, T. J. Watson Research Center, Yorktown Heights, NY 10598, USA}
\author{Julia E. Rice}
\affiliation{IBM Quantum, IBM Research - Almaden, 650 Harry Road, San Jose, CA 95120, USA}
\author{Tanvi P. Gujarati}
\affiliation{IBM Quantum, IBM Research - Almaden, 650 Harry Road, San Jose, CA 95120, USA}
\author{Sofia Gil}
\affiliation{Cornell University, Ithaca, NY 14850, United States}
\author{Gavin O. Jones*}
\affiliation{IBM Quantum, IBM Research - Almaden, 650 Harry Road, San Jose, CA 95120, USA}

\begin{abstract}
The simulation of chemical reactions is an anticipated application of quantum computers.
Using a Diels-Alder reaction as a test case, in this study we explore the potential applications of quantum algorithms and hardware in investigating chemical reactions. Our specific goal is to calculate the activation barrier of a reaction between ethylene and cyclopentadiene forming a transition state. To achieve this goal, we use quantum algorithms for near-term quantum hardware (entanglement forging and quantum subspace expansion) and classical post-processing (many-body perturbation theory) in concert. We conduct simulations on IBM quantum hardware using up to 8 qubits, and compute accurate activation barriers in the reaction between cyclopentadiene and ethylene by accounting for both static and dynamic electronic correlation. This work illustrates a hybrid quantum-classical computational workflow to study chemical reactions on near-term quantum devices, showcasing the potential of quantum algorithms and hardware in accurately calculating activation barriers.
\end{abstract}

\maketitle

\section{Introduction}

The Diels-Alder reaction, discovered by Otto Diels and Kurt Alder in 1928, remains a fundamental and extensively studied transformation in organic chemistry \cite{diels1928synthesen, woodward1969conservation, sauer1980mechanistic, houk1992transition, houk2021evolution}. The synthetic versatility of the Diels-Alder reaction is evident in its widespread use for the construction of complex natural products \cite{takao2005recent, juhl2009recent, nicolaou2002diels} and the design of novel materials \cite {franc2009diels, sanyal2010diels, gregoritza2015diels, munirasu2010functionalization, ratwani2023self, vauthier2019interfacial}. This reaction occurs between a conjugated diene and an alkene, referred to as a dienophile, and produces a cyclic compound, typically a six-membered ring. The reaction's efficiency and precise control over stereochemistry have established it as an indispensable tool for organic chemists seeking streamlined routes to elaborate molecular structures \cite {mcleod2019expanding, barber2018diels}. The extensive applicability of the Diels-Alder reaction in organic synthesis, combined with its intricate mechanistic aspects, positions it as a focal point for ongoing investigation and innovative advancements \cite{chauhan2022regioselectivity, xu2021enantioselective, zholdassov2023acceleration, chen2018ambimodal}.

The widespread importance and unique challenges of the Diels-Alder reaction make it a valuable testbed for near-term quantum computing algorithms \cite{kassal2011simulating,  motta2022emerging, cao2019quantum, bauer2020quantum, mcardle2020quantum} and hardware.
First, breaking and formation of bonds in the course of the reaction may lead to electronic wavefunctions with a multireference character, which can be captured to zeroth order by accurate active-space calculations. The energetics of the reaction then arise from a complex interplay between static and dynamic electronic correlation, the latter resulting from electronic transitions outside the active space. 
Finally, the reactivity and selectivity of the Diels-Alder reactions hinge on the characteristics of their transition states, which are typically more sensitive to approximations in the solution of the Schr\"{o}dinger equation than reactants and products, due to the presence of partial bonds. Therefore, accurate calculations of the Diels-Alder reaction pose a substantial challenge to quantum computing algorithms, since they require accounting for both static and dynamic electronic correlation in reactants, products, and transition states.

\begin{figure*}[t]
\includegraphics[width=0.8\textwidth]{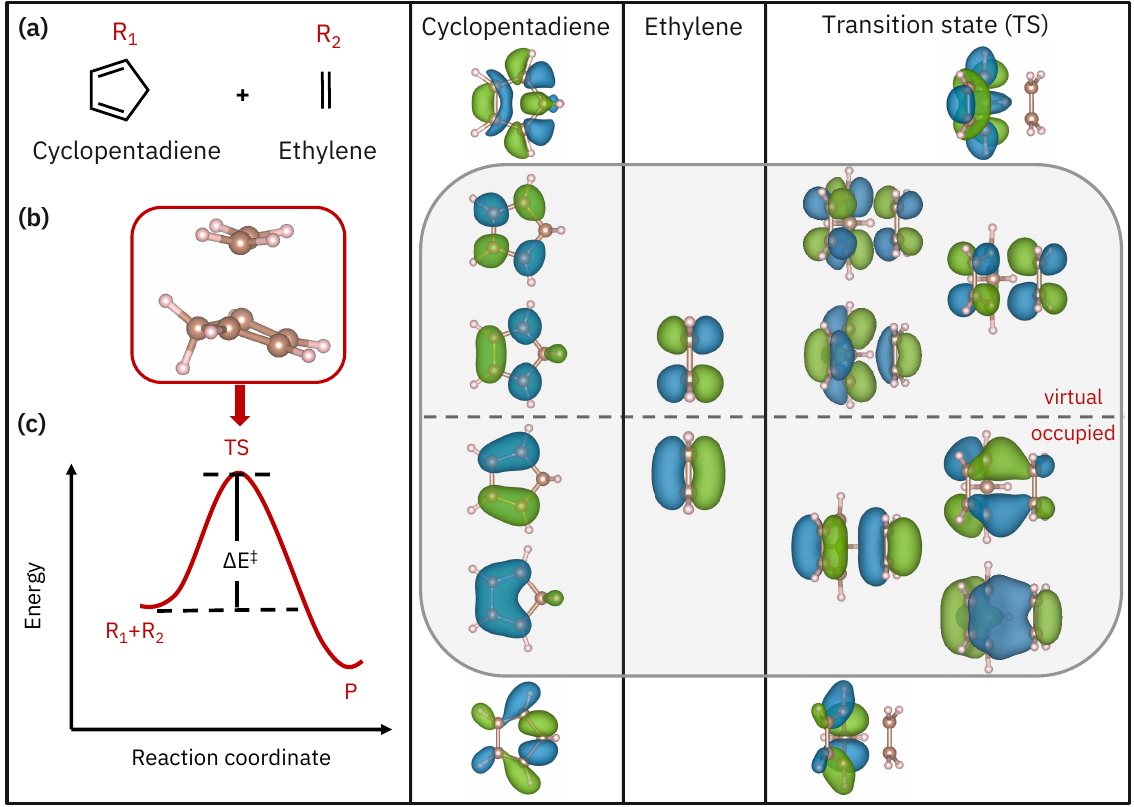}
\caption{Left panel: schematics of (a) the reactant molecules, (b) the transition state, and (c) the activation barrier denoted by $\Delta{E}^{\ddagger}$. The right three panels show the active-space orbitals for both reactants and the transition state (MP2 natural orbitals). A grey box highlights the AS(6e,6o) $\pi$ space of the reactants and transition state.}
\label{figure:IntroFigure}
\end{figure*}

In this work, we study the prototypical example of a Diels-Alder reaction, between cyclopentadiene and ethylene reacting in a synchronous ``aromatic type'' fashion where the reorganization of $\pi$ bonds in cyclopentadiene and ethylene (Figure \ref{figure:IntroFigure}) during bond formation leads to a bridged six-membered ring compound \cite{levandowski2015theoretical}. We explore the significance of this Diels-Alder reaction as a compelling testbed for the validation, combination, and refinement of quantum computing algorithms for near-term quantum devices. To that end, we employ hybrid quantum-classical algorithms to solve the Schr\"{o}dinger equation in an active space on a quantum computer, and then recover dynamical electronic correlation through classical post-processing. For active-space simulations, we use a qubit-reduction technique~\cite{mishmash2023hierarchical, eddins2022doubling, huembeli2022entanglement, setia2020reducing} called entanglement forging (EF)~\cite{eddins2022doubling, motta2023quantum, castellanos2023quantum} to define a variational ansatz in the context of the variational quantum eigensolver (VQE) method \cite{tilly2022variational, fedorov2022vqe}. To improve the quality of active-space simulations beyond the level of accuracy afforded by EF, we use a quantum subspace expansion (QSE) \cite{urbanek2020chemistry, motta2023subspace} based on single and double electronic excitations from the EF wavefunction.
For recovering dynamical electronic correlation, we integrate EF and QSE with second-order perturbation theory (PT2) \cite{angeli2001n, tammaro2023n}. We demonstrate the proposed algorithmic workflow (active-space calculations on quantum computers and classical post-processing exemplified by perturbation theory to recover dynamical electronic correlation) on classical simulators and quantum hardware, using up to 8 qubits and error mitigation techniques~\cite{nation2021scalable, pokharel2018demonstration, van2023probabilistic} to compute the activation energy of the Diels-Alder reaction.

The structure of this work is as follows: first, we detail the methods employed, emphasizing simulations on quantum hardware, including error mitigation techniques and measurement optimization. We then present and discuss results for predictions of the activation energy of the reaction on quantum simulators and quantum devices. The supplementary information includes additional details of the workflow used in this study.

\section{Methods}

\subsection{Active-space Selection}

In the Diels-Alder reaction, active electrons are defined as those participating in the breaking/formation of bonds as the reaction unfolds. The orbitals involved in the reaction involve two $\pi$ bonds contributed by the diene reacting with one $\pi$ bond contributed by the ene counterpart. These undergo conversion into partial $\pi$ bonds in the transition state, prior to the formation of two $\sigma$ bonds and one $\pi$ bond in the product. Overall, this process involves a 6 electron, 6 orbital active space (here denoted AS(6e, 6o)). Through the work of Houk and co-workers \cite{wilsey1999ground} on the retro reaction of norbornene breaking into cyclopentadiene and ethylene, it is known that the breaking/formation of bonds in a Diels-Alder reaction can be studied in an active space of 8 electrons in 8 orbitals, AS(8e,8o). Therefore, in this work, we simulate AS(8e,8o) of MP2 natural orbitals around the Highest Occupied Natural Orbital-Lowest Unoccupied Natural Orbital (HONO-LUNO) frontier \cite{gujarati2023quantum}, as shown in Figure~\ref{figure:IntroFigure}. We note that most of the active-space orbitals of the transition state have reactant-like character, consistent with the fact that the transition state has an ``early'' nature.
In particular, within the AS(8e,8o) active space shown in Figure~\ref{figure:IntroFigure}, one can recognize an AS(6e,6o) (enclosed in a gray box), spanned by the $\pi$ and $\pi^*$ orbitals of the reactants and transition state, respectively. In the remainder of this work, we therefore study the AS(6e,6o) alongside the larger AS(8e,8o) active space.

\begin{figure}[t!]
\includegraphics[width=0.5\textwidth]{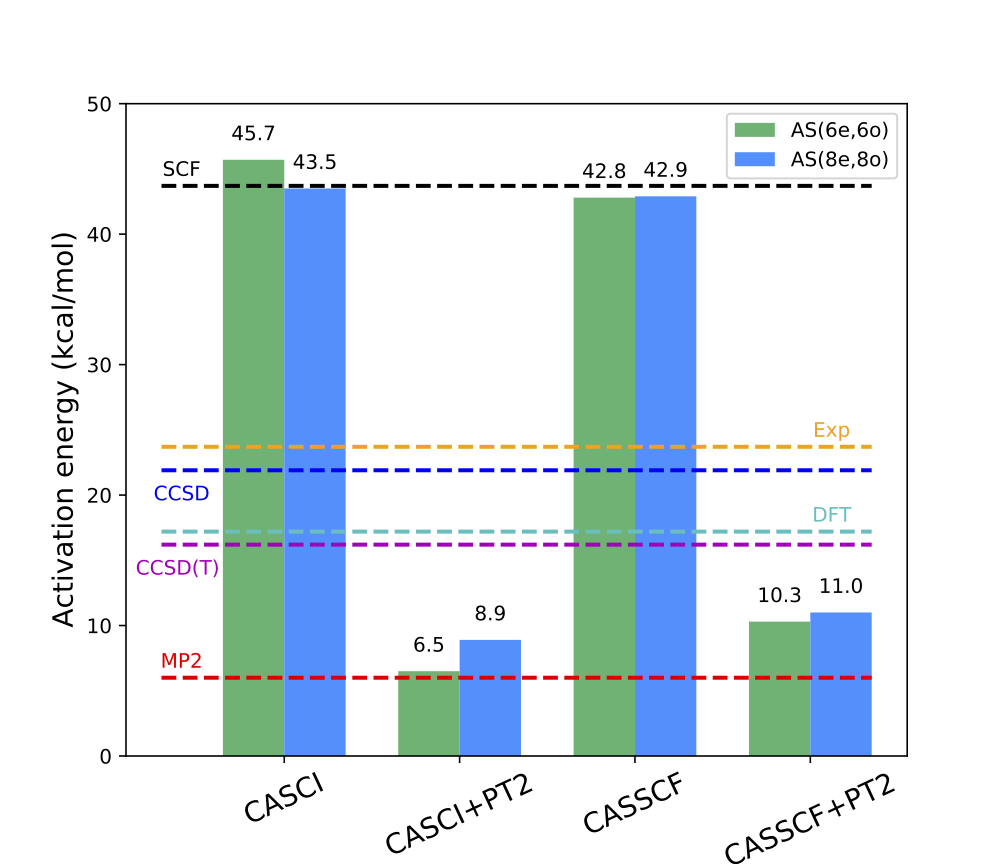}
\caption{Classical computational results for AS(6e,6o) (light green), AS(8e,8o) (dark blue), and various single-reference methods with differing treatments of electronic correlation (SCF, MP2, DFT with M06-2X functional, CCSD, and CCSD(T)), using the aug-cc-pVTZ basis set. Active-space methods included CASCI with natural orbitals, CASCI with natural orbitals combined with PT2 corrections to account for dynamical correlation, as well as CASSCF and CASSCF with PT2 corrections. The inclusion of dynamical correlation is essential to obtain realistic results.
}
\label{figure:classical}
\end{figure}

\subsection{Classical Methods}

We conducted classical electronic structure simulations using an aug-cc-pVTZ basis set \cite{woon1993gaussian} with PySCF \cite{sun2018pyscf, sun2020recent}. We obtained initial coordinates for the reactants and transition state from the prior study by Levandowski et al. \cite{levandowski2015theoretical}. 
Our CASCI calculations were performed using active spaces comprising MP2 natural orbitals. The CASCI energies for the active spaces of AS(6e,6o) and AS(8e,8o) were 45.7 kcal/mol and 43.5 kcal/mol, respectively. These values were compared to those obtained in previous computational studies \cite{guner2003standard, jorgensen1993ab} and to the experimentally determined (Exp) activation barrier \cite{roquitte1965thermal} of \(23.7 \pm 1.6\) kcal/mol in the gas phase (521-570 K) (Figure \ref{figure:classical}). The difference between the active-space energies and the experimental values underscores the significance of accounting for dynamical correlation to align the theoretical results with the experimental data. Consequently, we conducted second-order perturbation theory calculations to incorporate dynamical correlation. The activation barriers were found to be 6.5 kcal/mol and 8.9 kcal/mol for the active spaces of AS(6e,6o) and AS(8e,8o), respectively. CASSCF and CASSCF+PT2 calculations followed a similar trend. These classical electronic structure calculations serve as a reference point for evaluating the accuracy and precision of the quantum-classical algorithms employed in this study. Notably, the method that gets the closest to the experimental values is CCSD which suggests that the system under study is adequately described by a single reference wavefunction with dynamical correlation effects.

\subsection{Quantum Algorithms Overview}

\begin{figure}[t!]
\centering
\includegraphics[width=0.35\textwidth]{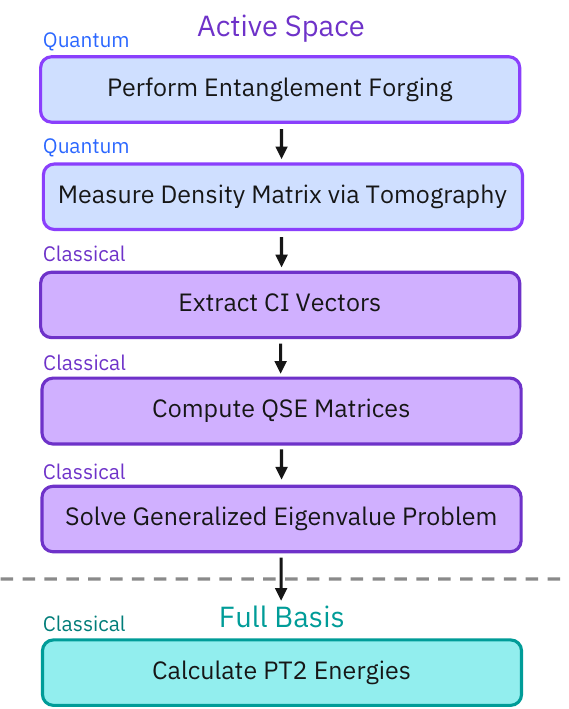}
\caption{Schematic representation of the hybrid approach workflow. Active-space calculations are performed on a quantum computer, followed by classical post-processing. After conducting an active-space variational calculation with entanglement forging (EF), we conduct an active-space quantum subspace expansion calculation to refine the EF results and second-order perturbation theory (PT2) to account for dynamical electron correlation.
To quantify and mitigate the errors occurring on quantum devices, we extract the density operators of the quantum states prepared by the device using tomography. We then project the density operator in the subspace of the Hilbert space with appropriate particle number and spin. We use the resulting projection (referred to as a CI vector) as a starting point for the QSE and PT2 calculations in lieu of the original density operator.
}
\label{figure:workflow}
\end{figure}

The workflow illustrated in Figure \ref{figure:workflow} uses a hybrid quantum-classical approach. First, we carry out active-space calculations using the VQE method, on quantum hardware, using the EF method to formulate a variational ansatz and reduce the number of qubits from $2N$ to $N$, where $N$ is the number of active-space orbitals. We then extract the density matrix of our system through tomographic measurements of the ground state on the quantum computer and we project it into the Hilbert space with the correct particle and spin number. The obtained projection (denoted as the CI vector in Figure \ref{figure:workflow}) is used as a starting point for QSE and PT2. This approach allows us to improve the quality of EF ground-state results and mitigate errors from the quantum device.

\subsubsection{Entanglement Forging}

Entanglement Forging (EF) is a qubit reduction technique that enables the simulation of electronic systems using only half the qubits required by a conventional simulation in the Jordan-Wigner representation, by mapping a spatial orbital to a single qubit instead of two. EF reduces the number of qubits by separately simulating electrons of opposite spins, and accounting for the correlation between opposite-spin electrons with classical post-processing based on a finite set of electronic configurations (bitstrings). EF was first demonstrated for the simulation of the water molecule \cite{eddins2022doubling} and later applied to study the excited-state dissociation of the sulfonium cation \cite{motta2023quantum}, as well as excitations in aromatic heterocycles \cite{castellanos2023quantum}. EF involves two steps: first, the identification of a subset of bitstrings to establish an initial multiconfiguration approximation of the electronic ground state; second, the selection of an appropriate quantum circuit. 

In the EF algorithm, the active-space Hamiltonian is expressed as a linear combination of tensor products,
\begin{equation}
H = \sum\limits_{\mu} \hat{A}_\mu \otimes \hat{B}_\mu,
\end{equation}
where $\hat{A}_\mu$ and $\hat{B}_\mu$ act on $\alpha$ and $\beta$ spin-orbitals, respectively.
The target wavefunction is represented by a Schmidt decomposition,
\begin{equation}
|\Psi_{\theta}\rangle = \sum\limits_{k}\lambda_k \hat{U}(\theta)|x_k\rangle \otimes \hat{U}(\theta)|x_k\rangle,
\end{equation}
in which the operator $\hat{U}(\theta)$ is a parameterized unitary, $\lambda_k$ is a set of Schmidt coefficients, and $|x_k\rangle$ are qubit computational-basis states represented by bitstrings. 

\begin{figure}[t!]
\includegraphics[width=0.45\textwidth]{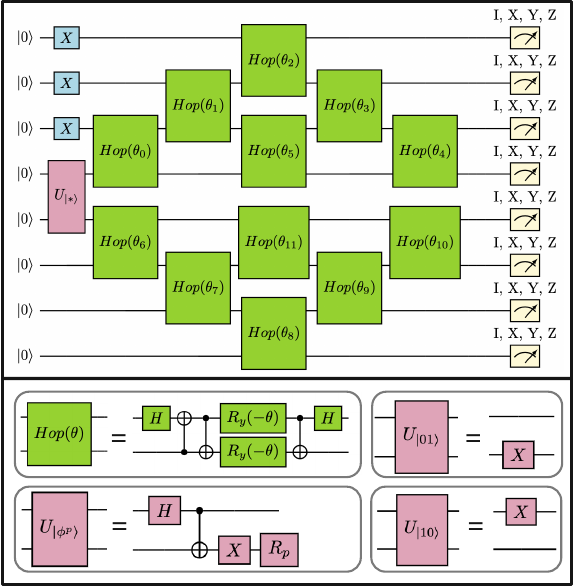}
\caption{Top: 8-qubit quantum circuit corresponding to the TS in AS(8e,8o) with state initialization run on a 27-qubit \textit{ibm\_auckland} device. A brick-wall arrangement of hop-gates (green), and measurement of single-qubit Pauli operators X, Y, and Z. Bottom: The hop-gate is compiled into single-qubit and CNOT gates. Two-qubit unitaries (highlighted in pink) transform the initial state $|00\rangle$ into various states, such as $|10\rangle$, $|01\rangle$, and $|\phi_p\rangle$ for $p = 0, 1, 2, \text{ and } 3$, corresponding to single-qubit gates ($R_0 = \mathbf{I}$, $R_1 = ZS$, $R_2 = Z$, and $R_3 = S$).}
\label{figure:EF-Circuits}
\end{figure}

To approximate the ground-state energy of our system, we evaluate the expectation value of $\hat{H}$

\begin{equation} \label{eq:3}
\langle \Psi_{\theta} | \hat{H} | \Psi_{\theta} \rangle = \sum_{kl\mu} \lambda^*_k \lambda_l A_{kl\mu} B_{kl\mu}.
\end{equation}
In Equation~\eqref{eq:3}, the matrices \(A_{kl\mu}\) and \(B_{kl\mu}\) are defined as
\begin{equation}
A_{kl\mu} = \langle x_k | \hat{U}^\dagger(\theta) \hat{A}_\mu \hat{U}(\theta) | x_l \rangle,
\end{equation}
\begin{equation}
B_{kl\mu} = \langle x_k | \hat{U}^\dagger(\theta) \hat{B}_\mu \hat{U}(\theta) | x_l \rangle.
\end{equation}
The bra and ket states \(\langle x_k |\) and \(| x_l \rangle\) are computational basis states labelled by bitstrings. For $k=l$, $A_{kl\mu}$ and $B_{kl\mu}$ are expectation values, that can be easily measured on quantum hardware. For $k\neq l$, they can be written as linear combinations of expectation values,
\begin{equation} 
A_{kl\mu} = \sum_{p=0}^3 \frac{(-i)^p}{4} \langle \phi^{p}_{kl} | \hat{A}_\mu | \phi^{p}_{kl} \rangle,
\end{equation}
where the superposition states are 
\begin{equation} 
| \phi^{p}_{kl} \rangle = \frac{ | x_k \rangle + i^p | x_l \rangle}{\sqrt{2}}.
\end{equation}
Figure \ref{figure:EF-Circuits} illustrates the 8-qubit EF circuit used in this work. The quantum circuits used in this study comprised two-qubit ``hop-gates'' that are both hardware-efficient and preserve the particle number. In Figure \ref{figure:EF-Circuits} the ``hop-gates'' are organized in a brick-wall configuration.  Details of the other circuits run in this study can be found in Appendix \ref{sec:app_HW_details}.

\subsubsection{Quantum Subspace Expansion}

To improve the accuracy of the EF results, we used the QSE method \cite{motta2023subspace, colless2018computation, mcclean2017hybrid, takeshita2020increasing} by applying single and double electronic excitations to the wavefunction obtained from EF as follows,
\begin{equation} \label{eq:8}
|\Psi\rangle = \alpha |\psi_{\mathrm{EF}}\rangle + \beta_{ki} \hat{a}^\dagger_{k} \hat{a}_{i} |\psi_{\mathrm{EF}}\rangle + \gamma_{kibj} \hat{a}^\dagger_{k} \hat{a}^\dagger_{b} \hat{a}_{j} \hat{a}_{i} |\psi_{\mathrm{EF}}\rangle.
\end{equation}
The coefficients $\alpha$, $\beta_{ai}$, and $\gamma_{aibj}$ were optimized variationally. In Equation~\eqref{eq:8}, $\hat{a}^\dagger_{k}$/$\hat{a}_{i}$ are the creation and annihilation operators, respectively for an electron in an occupied/virtual spin-orbital $k/i$. 

In this work, we focused on single and double electronic excitations within the same set of orbitals used to describe the ground state. This flavor of QSE can be regarded as a multi-reference CISD method, where the wavefunction, prepared on a quantum device, is not a single Slater determinant but a correlated electronic state. We classically realize a variational subspace spanned by a set of quantum states $\{ \psi_I \}$ as $| \psi_I \rangle = \hat{O}_{I} | \psi_\mathrm{EF} \rangle$, where $\hat{O}_I \in \{ \mathbf{I} , \hat{a}^\dagger_{k} \hat{a}_{i}, \hat{a}^\dagger_{k} \hat{a}^\dagger_{b} \hat{a}_{j} \hat{a}_{i} \}$, which can be generated via additional measurements and post-processing. The Hamiltonian is then diagonalized within the new state space, by solving the generalized eigenvalue problem $Hc = Sc{E}$ and obtaining a variational estimate of the ground state energy. More specifically, obtaining the expansion coefficients $c$ requires computing the matrix elements 
\begin{equation}
%\begin{split}
\label{eq:qse_matrices}
H_{ij} = \langle \psi_\mathrm{EF}| \hat{O}^\dagger_I \hat{H} \hat{O}_J |\psi_\mathrm{EF}\rangle = \mathrm{Tr} [\hat{O}^\dagger_I \hat{H} \hat{O}_J \hat{\rho}], 
%\end{split}
\end{equation}

\begin{equation}
%\begin{split}
S_{ij} = \langle \psi_\mathrm{EF}| \hat{O}^\dagger_I \hat{O}_J |\psi_\mathrm{EF}\rangle = \mathrm{Tr}[\hat{O}^\dagger_I \hat{O}_J \hat{\rho}].
%\end{split}
\end{equation}

We employ a quantum device to compute the matrix elements $H_{ij}$ and $S_{ij}$ by measuring the operators $\hat{O}^\dagger_I \hat{H} \hat{O}_J$ and $\hat{O}^\dagger_I \hat{O}_J$ respectively. Following \cite{motta2023quantum}, we conduct quantum state tomography on the EF circuits by performing measurements on up to $n=8$ qubits in the $3^n$ eigenbases of X, Y, and Z Pauli operators. Through this operation, we obtain a Bloch vector $a_P = \mbox{Tr}[P \rho]$, where $P$ is an $n$-qubit Pauli operator, and use it to calculate the matrix elements $H_{ij}$ and $S_{ij}$. 
We then use a classical computer to solve the generalized eigenvalue equation $Hc = Sc{E}$ and obtain approximate eigenpairs. The benefit of this approach is that QSE integrates into the VQE without necessitating any modifications to the quantum circuit, at the cost of additional measurements. Notably, it does not increase the depth of the quantum circuit required for preparing $|\psi_\mathrm{EF}\rangle$. This characteristic is beneficial, especially for near-term quantum hardware subject to qubit coherence times and two-qubit gate errors.
We remark that quantum state tomography is not required to measure the operators in Eq.~\eqref{eq:qse_matrices}. However, it is necessary to implement the classical post-processing operations described in the forthcoming Subsection~\ref{sec:pt2}.

\subsection{Hardware Calculations}

All EF calculations were executed on the 27-qubit \textit{ibm\_auckland} device, using the Qiskit Runtime library to interface the code with quantum devices. Jobs consisting of 300 circuits, and 10,000 shots for each circuit, were submitted on quantum hardware. Readout \cite{nation2021scalable} and dynamical decoupling \cite{pokharel2018demonstration} error mitigation techniques were employed to reduce noise originating from readout and quantum gates, respectively. Particle number was conserved through CI vector projection. Additional details can be found in Appendix \ref{sec:app_HW_details}.

\subsection{Error-weighted Pearson Correlation Analysis}

Quantum chemistry experiments on near-term quantum computers demand extensive time and resource utilization, underscoring the importance of achieving optimization in both aspects without compromising result fidelity. To optimize circuit time while maintaining performance, we conducted an in-depth analysis of quantum state tomography experiments on multiple IBM Quantum processors. Our approach involved calculating the Pearson correlation coefficient between quantum hardware results and the ground truth statevector, varying the number of shots, or alternatively, the number of circuit repetitions. Understanding the optimal number of shots required to achieve high-fidelity results enables the optimization of time and resources on near-term quantum computers without compromising result fidelity.

To that end, we first simulated tomography circuits for ethylene, cyclopentadiene, and the transition state using Qiskit Aer, representing the resulting statevector as a binary string of zeroes and ones, serving as a ground truth vector. Next, we selected several 27-Qubit IBM Quantum Falcon processors, including \textit{ibm\_algiers}, \textit{ibm\_cairo}, and \textit{ibm\_hanoi}. The experiments were executed using Qiskit Runtime, employing an optimization level of 3, readout error mitigation, and dynamical decoupling techniques. The variable explored in this study was the number of shots used in each experimental instance. The number of shots was systematically varied, and for each experimental instance, a Bloch vector was computed along with its associated errors. Subsequently, for every shot value, the resulting Bloch vector from the hardware run was correlated using the error-weighted Pearson correlation coefficient \( r_{\text{weighted}} \), defined as follows:

\begin{equation} \label{eq:Pearson}
    r_{\text{weighted}} = \frac{\sum_{i=1}^{n} w_i (X_i - \bar{X}_{\text{weighted}})(Y_i - \bar{Y})}{\sqrt{\sum_{i=1}^{n} w_i (X_i - \bar{X}_{\text{weighted}})^2} \sqrt{\sum_{i=1}^{n} (Y_i - \bar{Y})^2}},
\end{equation}

In Equation \ref{eq:Pearson} \( X \) and \( Y \) represent two sets of data, namely simulator and hardware block vectors, \(w_i\) are weights, \( X_i \) is the \( i \)-th data point in \( X \), and \( Y_i \) is the \( i \)-th data point in \( Y \). The mean of \( Y \) (\( \bar{Y} \)) and weighted mean of \( X \) (\( X_{\text{weighted}} \)) are expressed as:

\begin{equation}
\begin{split}
     \bar{Y} = \frac{1}{n} \sum_{i=1}^{n} Y_i,
\end{split}
\end{equation}

\begin{equation} \label{eq:A3}
\begin{split}  
    \bar{X}_{\text{weighted}} = \frac{\sum_{i=1}^{n} w_i X_i}{\sum_{i=1}^{n} w_i}. 
\end{split}
\end{equation}

In Equation \ref{eq:A3}, the weights are calculated as follows:

\begin{equation}
    w_i = 1 - \epsilon_i^2,
\end{equation}

where \( w_i \) is the weight for the \( i \)-th data point in \( X \), and \( \epsilon_i \) is the associated error.

\subsection{Perturbation Theory}
\label{sec:pt2}

An active-space calculation, carried out with an accurate solver and in carefully selected active space, can capture static correlation but not dynamical correlation arising from electronic interactions involving electrons in the inactive orbitals. A way of accounting for dynamical correlation is to combine active-space quantum computation with classical post-processing on the full basis set. An example is complete active-space second-order perturbation theory (CASPT2). Within CASPT2, the Hamiltonian is written as a sum of two terms $\hat{H} = \hat{H}_D + \hat{V}$, where $\hat{H}_D$ is the Dyall Hamiltonian, i.e., the sum between the active-space Born-Oppenheimer Hamiltonian and the restriction of the Fock operator to the non-active space, and $\hat{V} = \hat{H} - \hat{H}_D$ is treated as a perturbation. The second-order energy contribution is
\begin{equation} 
\Delta E_{\text{PT2}} = - \sum_{\nu \neq 0} \frac{| \langle \Psi_\nu | \hat{V} | \Psi_0 \rangle |^2}{E_\nu - E_0},
\end{equation}
where $(\Psi_\nu, E_\nu)$ are the eigenpairs of the Dyall Hamiltonian, and $\nu = 0$ labels the ground state. Implementing the exact (or uncontracted) NEVPT2 has a combinatorial cost with active-space size due to the sum over the excited states. This limitation can be remedied using strongly-contracted NEVPT2, which requires high-order ground-state reduced density matrices (RDMs), or partially-contracted NEVPT2, which approximates the sum over the excited states \cite{tammaro2023n}.

Implementing CASPT2 based on quantum computing data, specifically tomographic measurements, poses two challenges: first, active-space simulations conducted in the Fock space may break particle number conservation and other symmetries due to device noise, with detrimental impact on the accuracy of the ground and excited electronic states; second, statistical uncertainties need to be propagated from active-space quantities to $\Delta E_{\text{PT2}}$, leading to imprecise results.
In this work, we (i) sample the ground-state density matrix using quantum-state tomography and subsequently extract a Configuration Interaction (CI) vector by normalizing the density matrix's row/column entries for each sample. The resulting CI vector approximates the QSE wavefunction but has an exact particle number and considerably reduced statistical uncertainties. We then (ii) use each sampled CI vector as the input of a conventional CASPT2 calculation, and finally (iii) average the resulting PT2 energies.

\begin{figure*}[t!]
\includegraphics[width=0.8\textwidth]{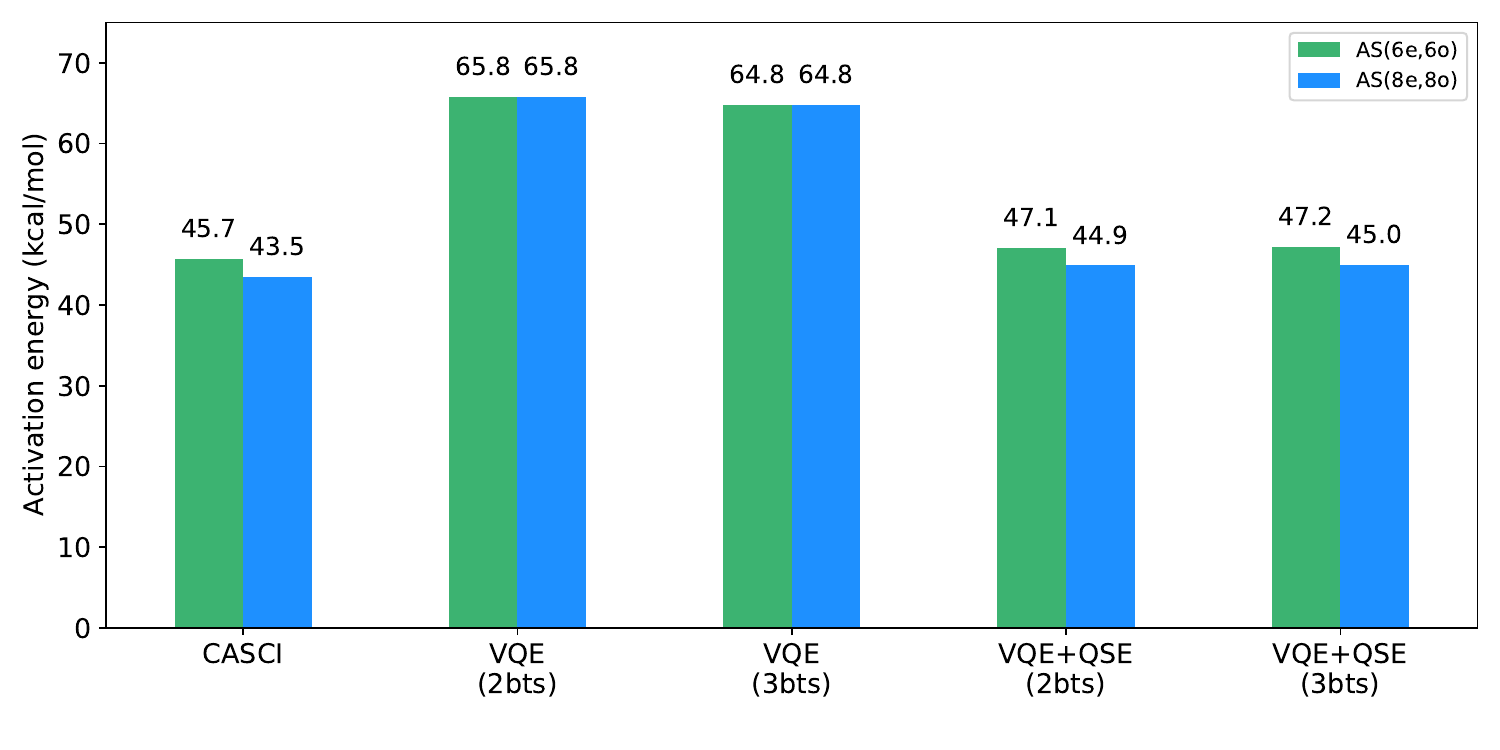}
\caption{Quantum simulations: VQE performed using the EF+QSE (denoted as VQE+QSE) for active-space AS(6e,6o) (green) and AS(8e,8o) (blue). Entanglement forging simulations were performed with 2 bitstrings (abbreviated as bts) for reactants and either 2 or 3 bitstrings for the transition state. The combination of entanglement forging and QSE results in a substantial reduction of approximately 10 kcal/mol in the activation barrier, closely aligning with outcomes obtained from classical CASCI calculations. Additionally, the introduction of an extra bitstring in entanglement forging for transition state calculations demonstrates minimal impact on the activation barrier.}
\label{figure:simulator}
\end{figure*}

\section{Results and Discussion}

\paragraph{Active-space quantum calculations}
In Figure~\ref{figure:simulator}, we present active-space calculations with EF and EF+QSE, carried out on classical simulators. In EF simulations, two bitstrings were used for the reactants and the transition state, respectively. These bitstrings, derived through a Schmidt decomposition of the FCI wavefunction, corresponded to the Hartree-Fock and HONO-LUNO excitation bitstrings. Given the relatively small size of the problem and the availability of the FCI solution, obtaining the FCI bitstrings and using them in our EF calculations allowed us to assess the effectiveness of the EF method. We further investigated the impact of a third bitstring for simulations of the transition state. Notably, simulated forged-VQE results for the transition state with 3 bitstrings resulted in lower energies, as shown in Figure \ref{figure:simulator}. This is expected because, as we include more bitstrings, we can achieve a more accurate representation of the electronic wavefunction. However, adding a third bitstring had a modest impact on the EF+QSE energy, which was approximately equal to the CASCI energy for both 2 and 3 bitstrings. Therefore, for computational efficiency, hardware calculations were executed using only two bitstrings for the transition state.
 
\begin{figure*}[t!]
\includegraphics[width=0.85\textwidth]{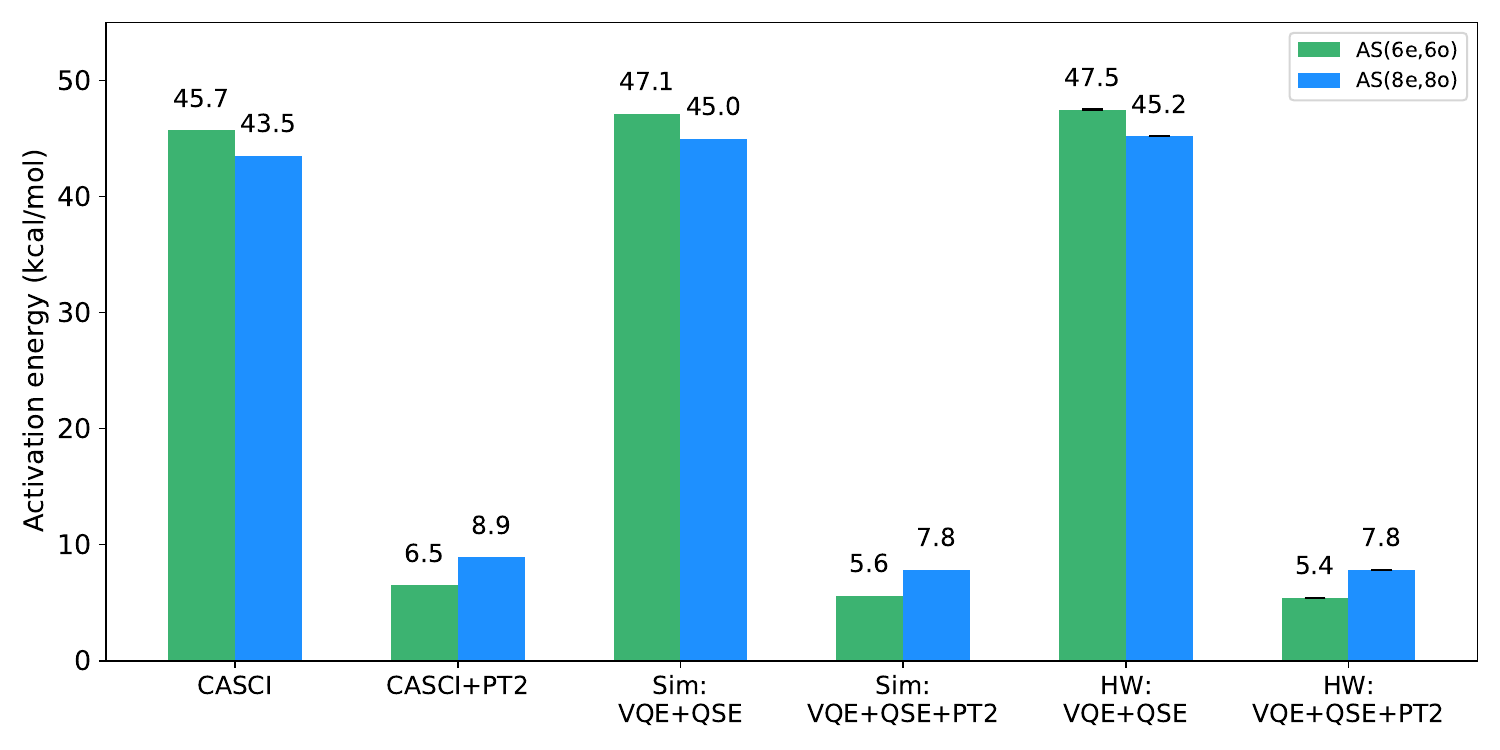}
\caption{Comparative analysis of classical CASCI and CASCI+PT2 calculations, quantum simulations (Sim), and hardware calculations (HW) for active spaces AS(6e,6o) (green) and AS(8e,8o) (blue) with and without second-order perturbation theory (PT2) for dynamical correlation. Quantum hardware results, employing error mitigation techniques, exhibit consistency with statevector simulations and classical CASCI calculations.}
\label{figure:pt2}
\end{figure*}

Hardware calculations, mitigated with readout and dynamical decoupling error suppression techniques, were further processed to ensure particle number preservation by extracting a CI vector representation as discussed in the previous Section. The results obtained in combination with QSE, closely matched the CASCI and statevector simulation results (Figure \ref{figure:pt2}). The associated statistical uncertainties for the activation barriers were 0.01 kcal/mol for both active spaces, with errors increasing with system size, but effectively cancelling out for the activation barriers. For example, when calculating absolute energies for ethylene AS(2e,2o), cyclopentadiene AS(4e,4o), and the transition state AS(6e,6o), respective errors were 0, $3.6 \times 10^{-6}$, and $1.6 \times 10^{-5}$ kcal/mol. Notably, the largest deviations in absolute energies were observed in the transition state, as expected due to its more complex electronic structure, compared to reactants.

\begin{figure*}[t!]
\includegraphics[width=0.8\textwidth]{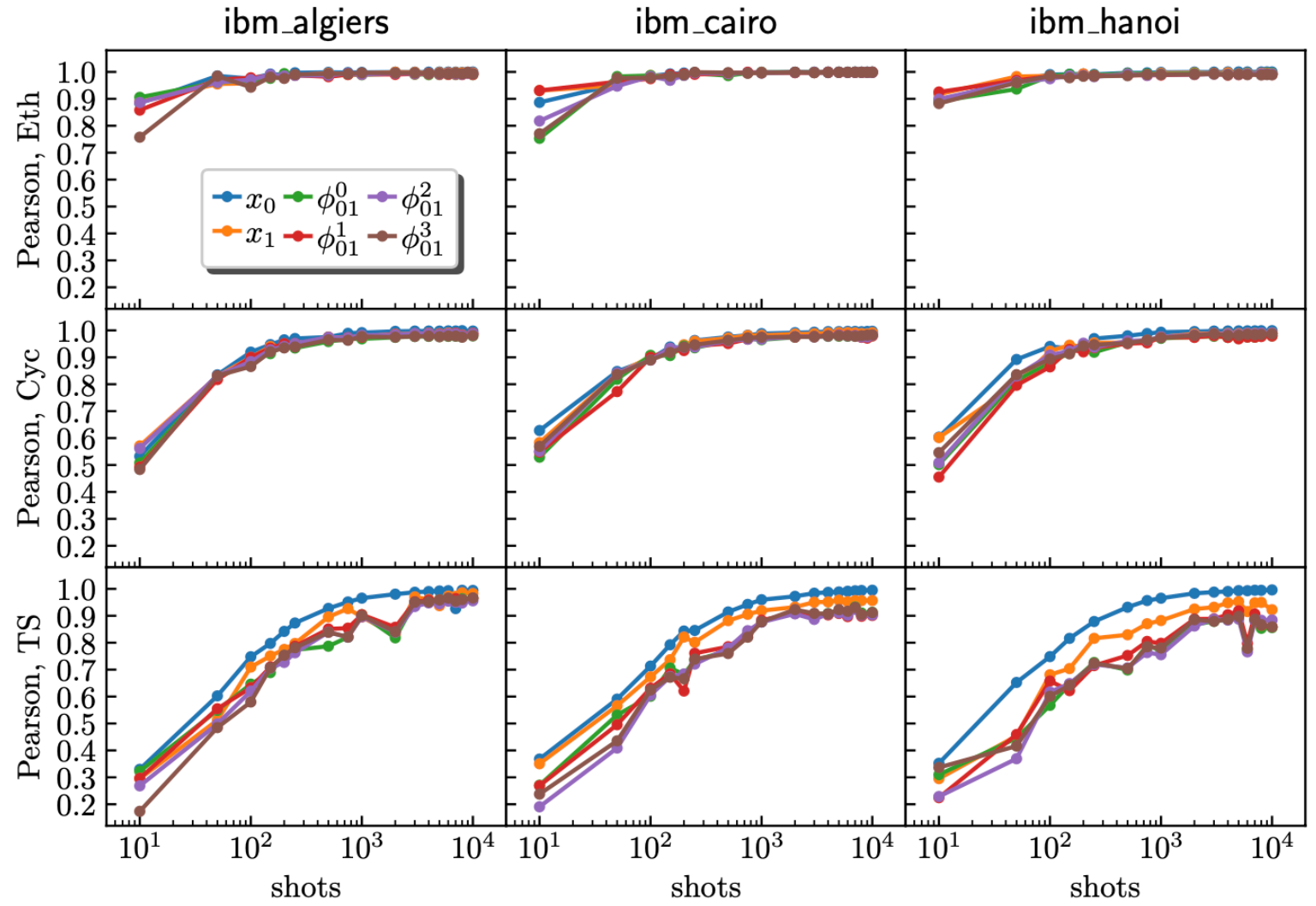}
\caption{Relationship between Pearson correlation (y-axis) and the number of shots (x-axis) for ethylene (Eth), cyclopentadiene (Cyc), and the transition state (TS) using data from \textit{ibm\_algiers}, \textit{ibm\_cairo}, and \textit{ibm\_hanoi} (from left to right). The data represent computational basis states ${\bf{x}}_0$ and ${\bf{x}}_1$, where ${\bf{x}}_0$ is the Hartree-Fock bitstring and ${\bf{x}}_1$ is a HONO-LUNO bitstring (e.g., for the transition state, ${\bf{x}}_k \in \{ |1111000 \rangle, |1110100 \rangle \}$). Superposition states $|\phi^p_{01} \rangle = \left( |{\bf{x}}_0 \rangle + i^p | {\bf{x}}_1 \rangle \right) / \sqrt{2}$ are marked as $\phi^p_{01}$, where $p=0,1,2,3$ respectively.}
\label{figure:pearson}
\end{figure*}

\paragraph{Pearson Correlation Shot Analysis}

Through additional optimization of hardware experiments and detailed Pearson correlation analysis \cite{yu2022inferential} for reactants and the TS, a trend emerged across the three quantum devices (\textit{ibm\_algiers}, \textit{ibm\_cairo}, and \textit{ibm\_hanoi}). The correlation between the hardware results and the ground truth statevector showed a significant improvement as the number of shots increased (Figures~\ref{figure:pearson}). However, our results also reveal a noticeable point of diminishing returns, and this point is contingent upon the circuit complexity inherent in the molecular system under investigation. For the TS, characterized by a high degree of circuit complexity, it became evident that even at the upper limit of 10,000 shots, there existed the potential for further enhancement in the quality of the result by gathering of additional shots. In contrast, for experiments involving cyclopentadiene with a moderate level of circuit complexity, an approximate shot count of 1,000 proved to be sufficient to reach a quality plateau. Notably, ethylene, which possesses the least complex circuit, achieved a plateau with only ~500 shots.

\paragraph{Perturbation theory}

After applying the PT2 correction to the activation barrier energies, the results were found to be consistent with classical CASPT2 energies, as illustrated in Figure \ref{figure:pt2}. The statistical uncertainties for the activation barriers were $3.7 \times 10^{-3}$ and $6.7 \times 10^{-3}$ kcal/mol for the AS(6e,6o) and AS(8e,8o), respectively. The low error bars are due to state projection, which reduces statistical fluctuations on the input of the PT2 calculation. Furthermore, the extraction of CI vectors from tomographic measurements yields a pure state (as opposed to a density operator) and ensures the correct number of electrons and spin. A detailed comparison between VQE and QSE, and VQE and QSE with CI vector purification in Table \ref{table:purification} shows that state purification significantly reduces errors.

\section{Conclusion}

In this study, we used the Diels-Alder reaction of cyclopentadiene with ethylene as a testbed for performing near-term simulations of reactions on quantum hardware. We computed the activation barrier of the reaction with an integrated combination of quantum algorithms for active-space calculations (Entanglement Forging, EF, and Quantum Subspace Expansion, QSE) and classical post-processing to recover dynamical electronic correlation (second-order perturbation theory, PT2). We demonstrated this computational workflow on classical simulators and quantum hardware, using up to 8 qubits and error mitigation. Additionally, insights derived from the Pearson correlation analysis enhanced our understanding of optimal shot selection and its impact on result fidelity in quantum experiments.

Our results pinpointed drastic approximations in the EF Ansatz, which overestimates the activation barrier by $\sim20$ kcal/mol compared to CASCI. We resolved the discrepancies between active-space quantum computing simulations with the chosen ansatz and CASCI by combining QSE with EF. However, CASCI (and any other active-space calculation) overestimates the activation energy, due to omission of dynamical electronic correlation. To overcome this limitation, we integrated EF and QSE with PT2, obtaining activation energies in agreement with CASPT2 within  $\sim1$ kcal/mol. These results, however, differ appreciably from experimental data and other classical calculations (CCSD, CCSD(T), DFT) due to the approximation of PT2.  

While our findings present compelling evidence of the effective application of the QSE method in refining ground-state approximations and enhancing the accuracy of VQE calculations, several important points should be noted about our methodology. Firstly, tomographic measurements are not scalable to larger system sizes. To address this issue, future work entails employing measurement optimization strategies such as Pauli grouping \cite{choi2023measurement} and cumulant approximation \cite{mazziotti1998approximate, zgid2009study}. This, in turn, would significantly reduce the computational demands associated with current implementations. Second, the classical CI vector sampling approach relies on classical representations of quantum information, thereby limiting its scalability. Additionally, among the methods that can be tested on Diels-Alder reactions in future research are: 
(i) embedding techniques~\cite{li2022toward, rossmannek2021quantum, gujarati2023quantum} to define active regions and correlate them with their environment, (ii) variational ansatzes to solve for the Schr\"{o}dinger equation in the active space seeking a balance between accuracy, computational cost, and hardware compatibility~\cite{d2023challenges, motta2023bridging, grimsley2019adaptive, ostaszewski2021reinforcement}, and (iii) approaches for recovering dynamical correlation, such as transcorrelated \cite{kong2012explicitly, motta2020quantum, kumar2022quantum}, downfolding \cite{huang2023leveraging}, and subspace methods \cite{takeshita2020increasing}. Our work highlights a Diels-Alder reaction as a compelling testbed for quantum algorithms and hardware, as it allows us to gauge their effectiveness in accounting for static and dynamical electron correlation in non-trivial situations (e.g. transition states), exposing and quantifying algorithmic approximations, and indicating areas and directions of improvement.

\appendix

\section{Entanglement Forging Hardware Calculations. Additional Details}
\label{sec:app_HW_details}

\begin{figure*}[t]
\includegraphics[width=0.85\textwidth]{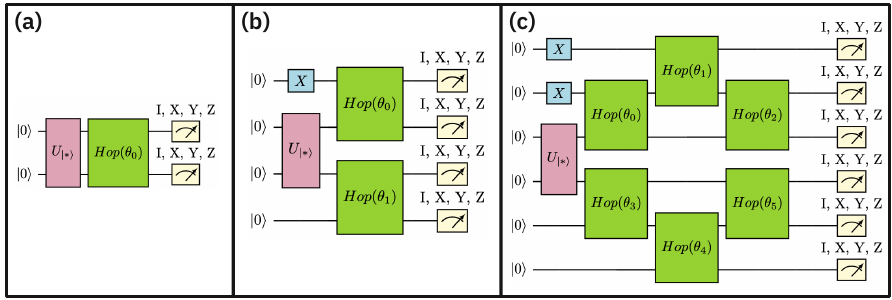}
\caption{Additional quantum circuits run in this study: (a) 2-qubit circuit for active space AS(2e,2o) for ethylene, (b) 4-qubit circuit for an active space AS(4e,4o) for cyclopentadiene, and (c) 6-qubit circuit representing active space AS(6e,6o) for cyclopentadiene and TS. The definitions of the two-qubit unitary ($U_{\ket{\star}}$) and the "hop-gate" are described in Fig. \ref{figure:EF-Circuits}.}
\label{figure:all-circuits}
\end{figure*}

Entanglement forging calculations on quantum hardware were performed for the reactants and transition state in the Diels-Alder reaction. The details on the number of qubits, parameters, gates, circuit depth and the total number of circuits are provided in Table \ref{table:hardware details}. The quantum circuits for the reactants and transition state are shown in Figures \ref{figure:EF-Circuits} and \ref{figure:all-circuits}.

\begin{table}[h!]
\centering
\begin{tabular}{cccccc}
\hline
System & Qubits & Parameters & Gates & Depth & Circuits \\
\hline
$\mathrm{C_2H_4}$ (2e,2o) & 2 & 3  & (12,4)  & 10 & 54  \\
$\mathrm{C_5H_6}$ (4e,4o) & 4 & 4  & (21,7)  & 10 & 486\\
$\mathrm{C_5H_6}$ (6e,6o) & 6 & 8  & (42,19) & 20 & 4374\\
$\mathrm{TS}$ (6e,6o)     & 6 & 8  & (42,19) & 20 & 4374 \\
$\mathrm{TS}$ (8e,8o)     & 8 & 14 & (72,37) & 30 & 39366\\
\hline
\end{tabular}
\caption{Key parameters in the study, including the number of qubits, variational parameters (one for every Hop gate and two for the Schmidt coefficients), the configuration of single- and two-qubit gates and the depth. Specifically, for each two-qubit unitary (denoted by $U_{\ket{\star}}$ in Fig. \ref{figure:EF-Circuits}), Hop gate, and measurement, there are 4, 4, 2 single-qubit gates and 1, 3, 0 two-qubit gates, respectively. The circuit depth is signifying the number of layers of quantum gates executed in parallel for computation completion. Tomography experiments on quantum hardware were run using Qiskit Runtime, which executes quantum circuits in sessions. Each Runtime job session on \textit{ibm\_auckland} contained a maximum of 300 circuits.}
\label{table:hardware details}
\end{table}

\section{CI Vector Purification}

We employed state projection and CI vector extraction as a noise mitigation/purification technique. The evaluation of noise in states by CI vectors significantly contributes to the reduction of error bars. These CI vectors represent pure states with the correct number of electrons and spin, enhancing the fidelity of our quantum computational analyses.\\

\begin{table}[h!]
\centering
\begin{tabular}{l@{\hspace{1.5em}}cc}
\hline
Method & $\Delta E^\ddagger$ (6e,6o) & $\Delta E^\ddagger$ (8e,8o) \\
\hline
SV (VQE+QSE) & 47.10 & 44.95 \\
HW (VQE+QSE) & 46.52 $\pm$ 2.10 & 44.49 $\pm$ 4.04 \\
HW (VQE+QSE+proj.) & 47.52 $\pm$ 0.01 & 45.24 $\pm$ 0.01 \\
HW (VQE+QSE+proj.+PT2) & 5.45 $\pm$ 0.004 & 7.85 $\pm$ 0.007 \\
\hline
\end{tabular}
\caption{ Comparison between
VQE+QSE,VQE+QSE with CI vector projection/purification, and VQE+QSE+PT2 with CI vector projection. Results were obtained using statevector (abbreviated SV), and \textit{ibm\_auckland} quantum hardware (abbreviated HW)} 
\label{table:purification}
\end{table}

\end{document}